\documentclass[nofootinbib,aps,prl,reprint,amsfonts,amssymb,amsmath]{revtex4-1}
\usepackage[bottom]{footmisc}
\usepackage{footnote}
\usepackage{perpage} 
\MakePerPage{footnote} 
\usepackage{hyperref}
\usepackage{todonotes}
\setuptodonotes{inline}

\DeclareMathOperator{\Tr}{Tr}
\usepackage{color}
\usepackage{comment}
\usepackage{mathtools}
\usepackage{multirow}
\usepackage{braket}

\begin{document}

\title{New Non-Supersymmetric Tachyon-Free Strings}

\author{Zihni Kaan Baykara$^1$, Houri-Christina Tarazi{$^{2,3}$}, Cumrun Vafa$^1$}

\affiliation{$^1$Department of Physics, Harvard University, Cambridge, Massachusetts, USA}

\affiliation{$^2$Enrico Fermi Institute \& Kadanoff Center for Theoretical Physics, University of Chicago, Chicago, IL 60637, USA}
\vspace{.2cm}
\affiliation{ $^3$Kavli Institute for Cosmological Physics, University of Chicago, Chicago, IL 60637, USA}

\begin{abstract}
In four decades of string theory research, only a handful of non-supersymmetric tachyon-free strings with only one neutral scalar at tree level were found. We construct new non-supersymmetric tachyon-free string theories using asymmetric orbifolds that serve as the lower-dimensional counterparts to the $O(16) \times O(16)$ string in 4d, 6d, and 8d, each featuring only one neutral scalar at tree level, chiral matter and positive leading order cosmological constant.  The 4d construction uses a quasicrystalline orbifold.
\end{abstract}

\maketitle

\section{Introduction}
The goal of any theory of quantum gravity is to provide realistic models that could describe both our cosmological history and the standard model physics. In particular, one would aim for non-supersymmetric vacua that have no moduli in the low energies, positive dark energy, chiral matter, and the standard model gauge group.

A particularly difficult endeavor in string theory is finding non-supersymmetric tachyon-free stable vacua. In the past decades of search, only three non-supersymmetric tachyon-free strings have been found in 10d: the $O(16)\times O(16)$ heterotic string \cite{Alvarez-Gaume:1986ghj, Dixon:1986iz}, the USp(32) string \cite{Sugimoto:1999tx} and the type 0'B string \cite{Sagnotti:1996qj,Sagnotti:1995ga}, with all having positive vacuum energy at leading order. The first two are almost rigid in the sense that their only massless scalar is the dilaton, but the USp(32) string also has a gravitino and realizes non-linear supersymmetry. Furthermore, the 0'B theory has two neutral scalars. Therefore, the $O(16)\times O(16)$ string is the only tachyon-free non-supersymmetric string with one neutral scalar and no gravitinos. This feature persists for the toroidal compactifications of the $O(16)\times O(16)$ string at special moduli with maximal non-abelian enhancement. However, the additional toroidal moduli will in general develop a mass or become tachyonic at one loop \cite{Ginsparg:1986wr}. There has been further study of the $O(16)\times O(16)$ vacua recently \cite{Fraiman:2023cpa} with new constructions including flux compactifications to $AdS_3 $ \cite{Baykara:2022cwj}.  

Finding more non-supersymmetric string theories is crucial for diversifying our non-supersymmetric string lab, which, until now, consisted of only three healthy models and their compactifications. Such strings should be tachyon-free in order to be  stable at tree level, and rigid, where the only neutral scalar is the string coupling, which is unavoidable in perturbative string theory. The $O(16)\times O(16)$ theory in particular is the only closed string example that has these features.
In this work, we provide the sought-after analogs of the $O(16)\times O(16)$ string directly in 4d, 6d, and 8d with no supersymmetry, no tachyons, one neutral massless scalar, together with additional massless charged fields at tree level.  At one loop the neutral scalar develops a running vacuum energy and the charged matter fields typically  pick up mass or become tachyonic.

Interestingly, all non-supersymmetric models with one neutral modulus have positive leading order potential. This is the case for all the models we present in this work as is for the  other three 10d string theories\footnote{In particular, the 0'B and $O(16)\times O(16)$ theory have more fermions than bosons and hence positive one-loop contribution. The USp(32) has a tree level positive contribution from the brane tension and a higher order negative contribution since it has more bosons. However, the latter theory has a spin-${3\over 2}$ field.}, including the 10d $O(16)\times O(16)$ string and its compactification on a circle at the point that there are no tachyons \cite{Fraiman:2023cpa}. 
This seems to suggest that any ``rigid" tachyon-free string theory necessarily has positive cosmological constant and chiral fermions providing a potential naturalness explanation for our positive vacuum energy and the existence of the Standard model fermions. 

The paper is organized as follows. In \autoref{sec:review} we review strings on orbifolds and set our notation. In particular, we will use asymmetric orbifolds \cite{Narain:1986qm} and the more special \textit{quasicrystalline} orbifolds \cite{Harvey:1987da}, recently studied in \cite{Baykara:quasi}. Details of the lattice constructions used are summarized in Appendix \ref{app:lattice}.
In \autoref{sec:4d} we describe the new non-supersymmetric tachyon-free rigid strings in 4d, 6d, and 8d obtained as asymmetric orbifolds of the heterotic string.  In Appendix \ref{app:constant} we present the 1-loop potentials computed. All the models have chiral fermions and are subject to anomaly cancellation. In Appendix \ref{app:anomalies} we summarize the anomaly cancellation methods used in each model.

\section{Orbifolds Review}\label{sec:review}
In this section we briefly review the important aspects of heterotic compactifications on tori and their orbifolds. 
\subsection{Torus compactifications}
Heterotic string theory compactified on a $d$-dimensional torus $T^d$ has ground state momenta $(p_L;p_R)$ taking values in an even self-dual lattice $\mathrm{\Gamma}^{d+16;d}\subset \mathbb R^{d+16;d}$, called the \textit{Narain lattice} \cite{Narain:1985jj, Narain:1986am}. Torus compactifications and background field data are completely characterized by the choice of the Narain lattice.

The mass spectrum is given by
\begin{align}
   & M^2_L=N_L+{p_L^2\over 2}-1 \\
    & M^2_R=N_R+{p_R^2\over 2}-\frac 1 2,
\end{align}
where $N_L,N_R$ are left and right moving oscillator numbers and $(p_L;p_R)\in \Gamma^{d+16;d}$ are the left and right ground state momenta in the internal dimensions.

\subsection{Lattice symmetries}
At special points in the moduli space, some automorphisms of the Narain lattice act as rotations on the left and right movers, without mixing the two. They form the \textit{Narain symmetry group}
\begin{align}
    \mathrm{Sym}(\Gamma^{d+16;d}):=\mathrm{Aut}(\Gamma^{d+16;d}) \cap \left(\mathrm{O}(d+16,\mathbb R)\times \mathrm{O}(d,\mathbb R)\right).
\end{align}
Such automorphisms act as symmetries on the worldsheet CFT, with T-duality at the self dual radius being the ur-example. Since these Narain symmetries $\theta$ are rotations, they have eigenvalues $e^{\pm 2\pi i \phi_L^i}$ on the left and $e^{\pm 2\pi i \phi_R^i}$ on the right. Therefore, we can equivalently characterize these rotations by their twist vectors
\begin{align}
    \phi = \left(\phi_L^1, \dots , \phi_L^{\frac d 2 +8} ; \phi_R^1,\dots , \phi_R^{\frac d 2}\right), \qquad \phi_L^i\in\mathbb Z,\quad\phi_{R}^i\in 2\mathbb Z,
\end{align}
where the $2\mathbb Z$ value specifies the $\mathrm{Spin}(d)$ uplift on the right.

A Narain symmetry $\theta$ is a \textit{quasicrystallographic symmetry} if it acts crystallographically on $\Gamma^{d+16,d}$ but not separately on the left and right \cite{Harvey:1987da}. In contrast, \textit{crystallographic symmetries} have separate crystallogaphic actions on the left and on the right. 

For example, there is no 2d lattice with a $\mathbb Z_5$ rotational symmetry due to the crystallographic restriction theorem. However, there is a $\mathbb Z_5$ action on the indefinite lattice $\Gamma^{2;2}_5$ constructed in Appendix \ref{app:lattice}, whose left and right movers form 2d quasicrystals with $\mathbb Z_5$ rotational symmetry. We present an in depth study of quasicrystallographic symmetries in \cite{Baykara:quasi}.

In addition to twists $\theta$, one can also act by a shift vector $v=(v_L;v_R)$ in the span of the Narain lattice $\Gamma^{d+16;d}$. The shift vector on the left and right movers are denoted by $v_L$ and $v_R$ respectively. We will present a shift vector $v$ on a lattice $\Lambda$ by its coefficients $(v^i)$ in the \textit{alpha basis}
\begin{align}
    v = v^i \alpha_i,\qquad v^i\in \mathbb Q
\end{align}
where the quadratic form of the lattice $\Lambda$ is given by the Gram matrix
\begin{align}
    G_\Lambda = \begin{pmatrix}
        \alpha_1\cdot \alpha_1  &\dots & \alpha_1\cdot \alpha_d\\
        \vdots & \ddots & \\
        \alpha_d \cdot \alpha_1 & \dots & \alpha_d\cdot \alpha_d
    \end{pmatrix}.
\end{align}
For a root lattice $\Lambda = \Gamma(\mathfrak g)$, the Gram matrix $G_\Lambda$ is the Cartan matrix and the alpha basis elements $\alpha_i$ are given by the simple roots.

The combined action of a twist and shift $ g=(\theta_L,v_L;\theta_R,v_R)$ lifts to an action $\hat{g}$ on the worldsheet states as follows:
\begin{align}
    \hat{g}\cdot \ket{p_L;p_R} &= e^{2\pi i (-p_L\cdot v_L + p_R \cdot v_R)} \ket{\theta_L\cdot p_L;\theta_R\cdot p_R},\\
    \hat g \cdot \alpha_{-k}^i \ket{0} &= e^{2\pi i \phi_L^i} \alpha_{-k}^i\ket{0},\\
    \hat g \cdot \bar \alpha_{-k}^i \ket{0} &= e^{2\pi i\phi_R^i} \bar \alpha_{-k}^i\ket{0},\\
    \hat g \cdot \ket{s_1,s_2,s_3,s_4} &= e^{2\pi i\phi_R\cdot s}\ket{s_1,s_2,s_3,s_4},\quad s_i=\pm \frac 1 2.
\end{align}
Here, the ground state momenta live in the Narain lattice $(p_L;p_R)\in \Gamma^{d+16;d}$, and $\alpha_{-k}^i,\bar \alpha_{-k}^i$ are complexified worldsheet oscillators on the left and right, and $\ket{s_1,s_2,s_3,s_4}$ is a Ramond ground state.

\subsection{Orbifolds}
String theory on orbifold backgrounds was introduced in \cite{Dixon:1985jw}. Geometrically, \textit{orbifolds} can be constructed from a torus $T^d$ by quotienting by a cyclic group $\mathbb Z_N=\langle g \rangle$ generated by isometry $g$ as $T^d/\mathbb Z_n$. On the worldsheet, the orbifolding procedure amounts to relaxing the boundary conditions of the string
\begin{align}
    X^i(\tau,\sigma+2\pi) = g^n\cdot X^i(\tau,\sigma) = \theta X^i(\tau,\sigma) +v^i,
\end{align}
and then projecting to the invariant subspace of $\hat{g}$
\begin{align}
    \hat{g}\cdot \ket{\psi} = \ket{\psi}.
\end{align}
The states with $g^n$-twisted boundary conditions make up the \textit{$\hat{g}^n$-twisted sector}. The $n=0$ sector corresponds to the \textit{untwisted sector}. Note that by CPT, $\hat{g}^n$ and $\hat{g}^{N-n}$-twisted sectors contain antiparticles of each other.

For geometric orbifolds, the action of $g$ on the Narain lattice is left-right symmetric as $\theta_L=\theta_R$ and $v_L=v_R$. Since the left and right degrees of freedom of strings are decoupled, one can generalize the orbifolding procedure to left-right asymmetric actions $g$ with $\theta_L\neq \theta_R$ or $v_L\neq v_R$. The procedure is carried out on the worldsheet CFT in a similar fashion by relaxing the left and right boundary conditions by $g=(\theta_L,v_L;\theta_R,v_R)$
\begin{align}
    X^i_{L,R}(\tau,\sigma+2\pi) = \theta_{L,R}\cdot X^i_{L,R}(\tau,\sigma)+v^i_{L,R},
\end{align}
and projecting to the invariant subspace of $\hat{g}$. Such orbifolds have no target space interpretation and are called \textit{asymmetric orbifolds} \cite{Narain:1986qm,Narain:1990mw}. But they can be related to geometric models as discussed in \cite{Kachru:1995wm,
Baykara:2023plc}.

Level matching is necessary and sufficient to ensure the consistency of the orbifolding procedure \cite{Vafa:1986wx}. In particular, for a $\mathbb Z_N$ orbifold, the energy levels on the left $E_L$ and right $E_R$ in each twisted sector must only differ by an integer multiple of $\frac 1 N$
\begin{align}
    E_R-E_L\in \frac{\mathbb Z}{N}.
\end{align}

We briefly describe the tools used to compute the mass spectrum in the twisted sectors. The ground state energy due to the twist in the $\hat{g}^n$-twisted sector is
\begin{align}
    (E_0)_{L}=\frac 1 2 \sum_i \{n\phi^i_{L}\} (1-\{n\phi^i_{L}\})
\end{align}
for bosonic, and
\begin{align}
    (E_0)_{R}=\frac 1 2 \sum_i \{n\phi^i_{L}\} -\frac 1 2
\end{align}
for fermionic sides,
where $0\leq \{a\}<1$ denotes the fractional part. In addition, the ground state energy due to surviving winding and momentum modes is
\begin{align}
    \frac{(p_{L,R}+nv^*)^2}{2}, \qquad (p_L;p_R)\in I^*,
\end{align}
where $I$ is the sublattice of $\Gamma^{16+d;d}$ that is fixed by the action $\theta^n$, $I^*$ is its dual lattice, and $v^*$ is the projection of the shift to $I$.

The oscillators in the $\hat{g}^n$-twisted sector are fractionally moded as $\alpha_{-k\pm n\phi^i}^i$. Lastly, the multiplicity of ground states is given by 
\begin{eqnarray}
    \chi(\theta^n)=\sqrt{det(1-\theta^n)\over |I^*/I|}.
\end{eqnarray}
 
\section{Tachyon-Free Orbifolds}\label{sec:4d}

In this section, we construct the lower dimensional analogs of the $O(16)\times O(16)$ heterotic string.

We have two criteria: a non-tachyonic spectrum, and the string coupling as the only massless neutral scalar at tree level. Note that the torus compactifications of the $O(16)\times O(16)$ string at enhanced gauge points fulfill these criteria---they are non-tachyonic and the only massless neutral scalar at tree level is the dilaton \cite{Ginsparg:1986wr}. Our main goal is to find heterotic orbifold examples beyond the $O(16)\times O(16)$ string paradigm.

There is no systematic way to ensure that a non-supersymmetric orbifold will have a non-tachyonic spectrum without carrying out the computations, but there is a way to ensure that there will be no neutral scalars. In particular, for geometric (i.e. symmetric) orbifolds $T^d/\mathbb Z_n$, there will always be neutral scalars associated with the moduli of the torus $T^{d}$. In contrast, asymmetric actions are only possible when the torus is string sized, therefore orbifolding at that point fixes the geometry and gets rid of neutral scalars.

To avoid tachyons on the other hand, one needs to get lucky---when there is a tachyonic left-mover state, there should be no level matching tachyonic right-mover, and vice versa. We have found such an example in each even dimension.\footnote{The ground state energies $(E_0)_L,(E_0)_R$ in 2d orbifolds are usually non-tachyonic, so it is significantly easier to find such non-supersymmetric models in 2d and we don't consider them here.} All the examples have chiral fermions, similar to the $O(16)\times O(16)$ string.

The reason we only have even dimensional examples is because a cyclic orbifold in odd dimensions would either have a fixed circle $S^1$ direction, hence an extra neutral scalar, or odd number of $-1$ eigenvalues in the twist, in which case the lift of the action to the worldsheet CFT would be more subtle \cite{Narain:1990mw,Harvey:2017rko}. It would be interesting if one can obtain similar strings in odd dimensions as well.
\subsection*{4d}

In this section we present a 4d non-supersymmetric heterotic orbifold with only one neutral complex scalar, given by the dilaton $\phi$ and the universal string axion $\star B$, i.e. the dual of the Kalb-Ramond field. All other scalars are charged and there are no tachyons in the twisted sectors. 

In fact, this model has an anomalous U(1) \cite{Alvarez-Gaume:1983ihn}, which becomes massive at 2-loops \cite{Dine:1987gj} and ``eats" the universal string axion to become its longitudinal mode. Details of the mechanism are summarized in \ref{app:4d}. As a result, the dilaton remains as the only neutral real scalar.

As is standard we present all Weyl fermions only as left handed, since a right handed Weyl fermion can equivalently be presented in terms of its left handed CP partner.\footnote{A right handed Weyl fermion in representation $R$ is equivalent to a left handed Weyl fermion in representation $\bar R$.}

The orbifold construction is as follows: we choose the Narain lattice $\Gamma(E_8) \oplus L\Gamma^{2;2}_5 \oplus \Gamma_5^{2;2}\Gamma_5^{2;2}$, where $L$ is a 8-dimensional lattice with $\det L=5$ and $\Gamma^{2;2}_5$ is a $\mathbb Z_5$ quasicrystal. Both of these lattices and their gluings are constructed explicitly in Appendix \ref{app:lattice}. 

For the orbifolding action, we twist by $\phi=(4,4,4,0^{8};2,2,2)/5$ and shift in the invariant lattice $\Gamma(E_8)\oplus L$ by
\begin{align}
v=(2,0,3,0,1,4,0,0,4,3,0,3,3,3,4,0)/5.
\end{align} 

The continuous gauge group is $\mathrm{SO}(10)\times\mathrm{SU}(5)\times \mathrm{SU}(3)\times \mathrm{SU}(2)\times \mathrm{U}(1)^4$.\footnote{It is interesting to note that quasicrystalline compactifications are not gauge enhanced points, therefore the rank of the orbifold gauge group is lower than what one would have expected from the usual crystallographic orbifolds.} The massless spectrum is
\begin{align}
    G + 84\times V+561\times F_c+9 \times F_0+354\times S_c+1\times  S_0,
\end{align}
where $G$ denotes the graviton, $S$ denotes a complex scalar, and $F$ denotes a left-handed Weyl fermion, and the subscripts $0$ and $c$ denote whether the state is neutral or charged with respect to the continuous gauge group. The representations for the scalars and fermions are given in \autoref{tab:nonsusy4d}.

The gauge group is obtained as follows. The invariant lattice has two orthogonal components $\Gamma(E_8)\oplus L$, whose corresponding gauge group is broken by the shift as
\begin{align}\label{eq:gaugebrok}
\begin{aligned}
    &E_8 \times (E_7\times U(1)) \\&\to (SO(10)\times SU(3)\times U(1)^3) \times (SU(5)\times SU(2)\times U(1)).
\end{aligned}
    \end{align}
The order of the $U(1)$ factors in \autoref{tab:nonsusy4d} is the same as in \eqref{eq:gaugebrok}. The Kac-Moody levels $k_i$ of the $U(1)$ factors are $30,100,60,300$ respectively, which we use to normalize the U(1) generators.

This model contains chiral charged fermions and hence anomalies can occur through non-trivial triangle diagrams with external gauge field/gravitons. All irreducible anomalies need to cancel identically but anomalies that can factorize as $X_2Y_4$ can be cancelled by the Green-Schwarz mechanism \cite{Green:1984sg} with a counterterm of the form $B\wedge X_2$.  This implies that the only fields that can be anomalous are U(1)s with $X_2\propto F_{U(1)}$. Such U(1)s are called ``anomalous" and occur when $\Tr(Q)\neq 0$ or when at least one of the external triangle legs contains that U(1). All such possibilities including the details of the GS mechanism in 4d are summarized in \autoref{app:4d}. The counterterm $B\wedge F_{U(1)}$
can be dualized to $\theta \partial_\mu A^\mu  $ where $\theta$ is the universal axion dual to the Kalb-Ramond field. However, such a coupling necessarily includes a mass term for the gauge field with $\theta $ becoming its longitudinal mode. The mass term for the gauge field is produced at 2-loops in string theory \cite{Dine:1987xk}. Additionally, a potential for the charged scalar fields will be generated at 1-loop \cite{Dine:1987gj} and for the dilaton at 2-loops \cite{Atick:1987qy}.

\begin{table}[h!]
\begin{align*}
    \begin{array}{|c|c|c|}
    \hline
         & \multicolumn{2}{|c|}{\mathrm{SO}(10)\times\mathrm{SU}(5)\times \mathrm{SU}(3)\times \mathrm{SU}(2)\times \mathrm{U}(1)^4\text{ reps}}\\
         \hline
                        \text{Sector}
                        &  \text{Complex scalars} & \text{Left handed Weyl fermions}\\
                        \hline
         \multirow{12}{*}{\text{Untwisted}}& (\mathbf{1,1,1,1})_{0,0,0,0} & 9(\mathbf{1,1,1,1})_{0,0,0,0}\\
                         & 3(\mathbf{16,1,1,1})_{0,0,0,-15} &(\mathbf{16,1,1,1})_{0,0,0,-15}\\
                         &   3(\mathbf{10,1,\bar 3,1})_{0,0,0,10} &(\mathbf{10,1,\bar 3,1})_{0,0,0,10}\\
                         &   3(\mathbf{1,5,1,1})_{-2,-10,-2,0} &(\mathbf{1,5,1,1})_{-2,-10,-2,0}\\
                         &   3(\mathbf{1,\overline{10},1,2})_{-1,5,-1,0} & (\mathbf{1,\overline{10},1,2})_{-1,5,-1,0}\\
                         &   3(\mathbf{1,1,1,2})_{5,-5,5,0} &(\mathbf{1,1,1,2})_{5,-5,5,0}\\
                         &  &3(\mathbf{16,1,3,1})_{0,0,0,5}\\
                        & &3(\mathbf{1,10,1,1})_{-4,0,-4,0}\\
                        & &3(\mathbf{1,1,\bar{3},1})_{0,0,0,-20}\\
                        & &3(\mathbf{1,5,1,2})_{3,5,3,0}\\
                         & &3(\mathbf{1,\bar 5,1,1})_{2,-10,2,0}\\
                         \hline
        \multirow{11}{*}{ $\hat{g}+\hat{g}^4$ } &   & 15(\mathbf{1,1,1,2})_{-1,-3,-1,12}\\
    &&15(\mathbf{1,1,1,1})_{-2,7,1,12}\\
    &&15(\mathbf{1,1,1,1})_{0,7,-3,12}\\
    &&5(\mathbf{1,1,1,2})_{1,-3,-5,12}\\
    &&5(\mathbf{1,1,1,2})_{-3,-3,3,12}\\
    &&5(\mathbf{1,5,1,1})_{2,2,2,12}\\
    &&5(\mathbf{1,\bar 5,1,1})_{2,-3,-1,12}\\
    &&5(\mathbf{1,\bar 5,1,1})_{0,-3,3,12}\\
    &&5(\mathbf{1,1,\bar 3,2})_{-1,-3,-1,-8}\\
    &&5(\mathbf{1,1,\bar 3,1})_{0,7,-3,-8}\\
    &&5(\mathbf{1,1,\bar 3,1})_{-2,7,1,-8}\\
    \hline
        \multirow{3}{*}{ $\hat{g}^2+\hat{g}^3$ } 
    & 15(\mathbf{1,1,\bar 3,1})_{4,-1,1,4}&5(\mathbf{1,1,\bar 3,1})_{4,-1,1,4}\\
    &15(\mathbf{1,1,\bar 3,1})_{2,-1,5,4}&5(\mathbf{1,1,\bar 3,1})_{2,-1,5,4}\\
    &15(\mathbf{1,1,\bar 3,1})_{-2,-6,-2,4}&5(\mathbf{1,1,\bar 3,1})_{-2,-6,-2,4}\\
        \hline
    \end{array}
\end{align*}
\caption{The charges of the scalars and fermions of the non-supersymmetric 4d quasicrystalline $\mathbb Z_5$ orbifold. The spectrum is given in terms of complex scalars and left-handed Weyl fermions.}
    \label{tab:nonsusy4d}
\end{table}

The leading order contribution to the canonically normalized dilaton potential in Einstein frame \footnote{The canonically normalized contribution from converting to Einstein frame is $e^{-D\over \sqrt{D-2}}\hat{\phi}$ which exactly multiplies all 1-loop corrections.} is 
\begin{eqnarray}
    V_{1-loop}(\hat{\phi})\approx e^{-2\sqrt{2}\hat{\phi}}\left(3.13 \times 10^{-2}\right)M_s^4,
\end{eqnarray}
which is positive.

Therefore, we have a very interesting example with Standard model like gauge symmetry and chiral matter. The model has only one neutral scalar, the dilaton\footnote{The dual to the Kalb-Ramond 2-form is an axion and hence a scalar field which will also have a potential with a  two-loop mass term and will become the longitudinal mode of the abelian U(1)(\autoref{app:4d}). }, for which there is a potential making it massive at higher order \autoref{app:4d} with a 2-loop  contribution to the potential due to the massive U(1).

We can also compute the beta function for these gauge groups to understand their IR behavior.
The computation of the one-loop beta function in \autoref{app:4d} shows that the $\mathrm{SO}(10)$ and $\mathrm{SU}(5)$ factors are confined and all other gauge factors are IR free. All higher dimensional models are automatically IR free due to the gauge kinetic terms being irrelevant.

\subsection*{6d}
We continue our search for tachyon-free models with one neutral scalar in higher dimensions.
The next example is in 6d. All other scalars are complex and charged, and the spectrum is tachyon-free. In 6d, CP preserves the chirality and therefore we indicate the chirality when writing the spectrum.

 Choose the Narain lattice $\Gamma(E_8\times E_8) \oplus \Gamma^{4;4}(A_4)$. Twist by $\phi=(0^{10};2,4)/5$ and shift in the invariant lattice $\Gamma(E_8\times E_8)\oplus \Gamma(A_4)$ by
\begin{align}
    v=(3,3,1,4,4,1,2,2,4,4,1,1,2,4,2,3,3,3,2,3)/5.
\end{align}

The gauge group corresponding to the invariant lattice $\Gamma(E_8\times E_8)\times \Gamma(A_4)$ is broken by the shift as
\begin{align}
    E_8 \times E_8 \times SU(5) \to SU(5)^2 \times SU(5)^2 \times U(1)^4.
\end{align}
So the continuous gauge group is $\mathrm{SU}(5)^4\times U(1)^4$. The Kac-Moody levels of the $U(1)$ factors are $4,60,10,30$, which we use to normalize the $U(1)$s to have level 1. The massless spectrum is
\begin{align}
    G+B+100\times V+460\times F_c+460\times \bar{F}_c\\+460\times S_c+1\times S_0.
\end{align}
Here, $G$ and $B$ denote the graviton and the Kalb-Ramond fields, $S_0$ denotes a real neutral scalar (the dilaton), $S_c$ denotes a complex and charged scalar, and $F_c, \bar F_c$ respectively denote charged left or right handed Weyl fermions. The representations for the charged spectrum are given in \autoref{tab:nonsusy6d}.

\begin{table}[h!]
\begin{align*}
    \begin{array}{|c|c|}
    \hline
         \text{Sector} & \multicolumn{1}{|c|}{\mathrm{SU}(5)\times\mathrm{SU}(5)\times \mathrm{SU}(5)\times \mathrm{SU}(5)\times \mathrm{U}(1)^4\text{ reps}}\\
                        \hline
         \multirow{14}{*}{\text{Untwisted}}
                         & ^R(\mathbf{10,5,1,1})_{0,0,0,0} \\
                        & ^R(\mathbf{\overline{5},10,1,1})_{0,0,0,0} \\
                        & ^L(\mathbf{1,1,10,5})_{0,0,0,0} \\
                        & ^L(\mathbf{1,1,5,\overline{10}})_{0,0,0,0} \\
                         &   ^R(\mathbf{1,1,1,1})_{-1,3,2,6} \\
                        &   ^R(\mathbf{1,1,1,1})_{0,8,-3,1} \\
                        &   ^R(\mathbf{1,1,1,1})_{1,-7,-3,1} \\
                        &   ^R(\mathbf{1,1,1,1})_{-2,-2,2,4} \\
                        &   ^R(\mathbf{1,1,1,1})_{2,-2,2,-4} \\
                        &   ^L(\mathbf{1,1,1,1})_{1,1,4,2} \\
                        &   ^L(\mathbf{1,1,1,1})_{0,-4,-1,7} \\
                        &   ^L(\mathbf{1,1,1,1})_{2,6,-1,-3} \\
                        &   ^L(\mathbf{1,1,1,1})_{-1,-9,-1,-3} \\
                        &   ^L(\mathbf{1,1,1,1})_{-2,6,-1,-3} \\
                         \hline
        \multirow{5}{*}{$\hat{g}+\hat{g}^4$} &  ^R(\mathbf{\overline{5},1,1,5})_{0,-4,0,-2}  \\
        &  ^R(\mathbf{\overline{5},1,1,5})_{0,0,2,0}  \\
        &  ^R(\mathbf{\overline{5},1,1,5})_{1,1,-1,-1}  \\
        &  ^R(\mathbf{\overline{5},1,1,5})_{-1,3,0,0}  \\
        &  ^R(\mathbf{\overline{5},1,1,5})_{0,0,-1,3}  \\
    \hline
        \multirow{5}{*}{$\hat{g}^2+\hat{g}^3$} &  ^L(\mathbf{1,5,5,1})_{1,1,1,-1}  \\
        &  ^L(\mathbf{1,5,5,1})_{0,-4,-1,1}  \\
    &  ^L(\mathbf{1,5,5,1})_{-1,-1,0,-2}  \\
    &  ^L(\mathbf{1,5,5,1})_{0,0,1,3}  \\
    &  ^L(\mathbf{1,5,5,1})_{0,4,-1,-1}  \\
        \hline
    \end{array}
\end{align*}
\caption{The charged spectrum of the non-supersymmetric 6d $\mathbb Z_5$ orbifold. For each representation listed here, there is a complex scalar and a symplectic Majorana-Weyl fermion of the indicated chirality.}
    \label{tab:nonsusy6d}
\end{table}

The anomaly polynomial is given by 
\begin{eqnarray}
 (2\pi)^4  I_{8}={1\over 12}X_4Y_4
\end{eqnarray}

with 
\begin{align}
\begin{split}
    X_4=&-15\sum_{a=1,2}\Tr(F^2_{SU(5)_a})+15\sum_{a=3,4}\Tr(F^2_{SU(5)_a})\\&+ \sum_{i\geq j}{1\over 4}{b_{ij}}{\Tr(F_{U(1)_i})\over\sqrt{k_i}}{\Tr(F_{U(1)_j})\over \sqrt{k_j}}
    \end{split}
\end{align}
where 
\begin{eqnarray}
    b_{ij}=\left(
\begin{array}{cccc}
 0 & 120 & 60 & 60 \\
 120 & 60 & 60 & -420 \\
 60 & 60 & -15 & 120 \\
 60 & -420 & 120 & 15 \\
\end{array}
\right)
\end{eqnarray}
and 
\begin{align}
 Y_4=  \sum_i\Tr (F^2_{U(1)_i})+2 \sum_a \Tr(F^2_a)-\Tr(R^2)
\end{align}
where we have rescaled the U(1)s to bring them in the canonical form $Y_4$.
The $I_{8}$ anomaly  is cancelled with the GS counterterm $ B\wedge X_4$.

The string frame cosmological constant for this theory is calculated in \autoref{app:constant} and gives the leading term in the dilaton potential in Einstein frame
\begin{eqnarray}
  V(\hat{\phi})_{\text{1-loop}}       \approx e^{-3\hat{\phi}}\left(2.89 \times 10^{-3}\right)M_s^{6},
\end{eqnarray}
which is positive at leading order.

\subsection*{8d}
Next we give a non-supersymmetric and tachyon-free 8d orbifold model. Similarly, the only neutral scalar is the dilaton with all other scalars charged. The situation of the fermions is similar to that in 4d, so we will present the spectrum only in terms of left handed Weyl fermions.

Consider the Narain lattice $\Gamma(E_8\times E_8)\oplus \Gamma^{2;2}(A_2)$ with a twist of the form $\phi=(0^{9};2/3)$ and a shift in the invariant lattice $\Gamma(E_8\times E_8)\oplus \Gamma(A_2)$ given by
\begin{align}
    v= (2,1,0,2,2,2,1,2,0,2,0,2,2,0,2,2,1,0)/3.
\end{align}
The gauge group corresponding to the invariant lattice $\Gamma(E_8\times E_8)\oplus \Gamma(A_2)$ is broken by the shift as
\begin{align}
    E_8 \times E_8 \times SU(3) \to SU(9) \times SU(9) \times U(1)^2.
\end{align}
The full continuous gauge group is $SU(9)^2 \times U(1)^2$. The levels of the $U(1)$ factors are $6,18$. The total massless spectrum is
\begin{align}
    G +B + 162\times V + 414\times F_c + 414\times S_c + 1\times S_0.
\end{align}
Here, $S_0$ denotes a real neutral scalar, $S_c$ denotes a complex and charged scalar, and $F_c$ denotes a charged left handed Weyl fermion. The representations for the charged spectrum are given in \autoref{tab:nonsusy8d}.

\begin{table}[h!]
\begin{align*}
    \begin{array}{|c|c|}
    \hline
         \text{Sector} & {\mathrm{SU}(9)\times\mathrm{SU}(9)\times \mathrm{U}(1)^2\text{ reps}}\\
                        \hline
         \multirow{5}{*}{\text{Untwisted}}
                         & (\mathbf{84,1})_{0,0} \\
                        & (\mathbf{1,{84}})_{0,0} \\
                        & (\mathbf{1,1})_{0,-6} \\
                        & (\mathbf{1,1})_{-3,3} \\
                         &   (\mathbf{1,1})_{3,3} \\
                         \hline
        \multirow{3}{*}{$\hat{g}+\hat{g}^2$} &  (\mathbf{{9},9})_{-1,1}  \\
                            &  (\mathbf{{9},9})_{1,1}  \\
                            &  (\mathbf{{9},9})_{0,-2}  \\
        \hline
    \end{array}
\end{align*}
\caption{The charged spectrum of the non-supersymmetric 8d $\mathbb Z_3$ orbifold. For each representation listed here, there is a complex scalar and a left handed Weyl fermion(together with its CP conjugate).}
    \label{tab:nonsusy8d}
\end{table}

Anomalies for this theory can be computed by 5-leg 1-loop diagrams with external gauge fields or gravitons. Such anomalies can be cancelled using the Green-Schwarz mechanism which is summarized in \autoref{app:8d}.
The anomaly polynomial is given by 
\begin{eqnarray}
 (2\pi)^5  I_{10}= {1\over 12}X_6Y_4
\end{eqnarray}
with
\begin{align}
\begin{split}
    X_6=&{-1\over 4}\Tr(F^3_{U(1)_2})+{3}  \Tr(F^3_a)+{3\over 4}  \Tr(F_{U(1)_2})\Tr(F_{U(1)_1}^2)
    \end{split}
\end{align}
and 
\begin{align}
 Y_4=  \Tr (F^2_{U(1)_2})+2 \Tr(F^2_a)-\Tr(R^2)+\Tr( F_{U(1)_1}^2)
\end{align}
where we have rescaled the U(1)s to bring them in the canonical form $Y_4$.
The $I_{10}$ anomaly  is cancelled with the GS counterterm $ B\wedge X_6$.

In 8d there can also be global anomalies as discussed in \cite{Alvarez-Gaume:1983ihn} but we have an even number of fermions and hence the model is anomaly free. Similarly, to the Swampland observation in \cite{Montero:2020icj} the rank is even too and it would be interesting to investigate if similar arguments can be applied for these cases. Similar, generalized anomalies have been computed in \cite{Basile:2023knk} for non-supersymmetric strings which could directly be applied in our models.

The global gauge anomaly \cite{Witten:1982fp,Elitzur:1984kr} is also absent since $\pi_8(SU(9))=0$ and the \cite{Lee:2022spd} is absent since we do not have the adjoint of $SU(9)$.

The string frame cosmological constant for this theory is computed by the 1-loop dilaton tadpole computed in \autoref{app:constant} which in Einstein frame gives the leading contribution to the dilaton potential:
\begin{align}
  V_{1-loop}(\hat{\phi}) \approx e^{{-8\over \sqrt{6}}\hat{\phi}} \left(1.26 \times 10^{-4}\right)M_s^{8},
\end{align}
which is positive. Therefore, all three models we have constructed have positive potential at leading order in 4,6 and 8 dimensions.

\section{Conclusion}
We have extended the landscape of non-supersymmetric, tachyon-free string theories by introducing novel theories in 4d, 6d, and 8d that mirror the characteristics of the $O(16) \times O(16)$ heterotic string. The only neutral scalar in each theory is the string coupling and each theory has a positive potential at leading order. It would be worthwhile to search for more such examples including in odd dimensions.
Other non-supersymmetric orbifold model searches include \cite{Font:2002pq,Sasada:1995wq,Avalos:2023mti,Anastasopoulos:2004ga}.

The theories we have constructed all come from orbifold models of the heterotic string. This implies that in 4d we only have one universal axion which sets the mass of the ``anomalous" U(1). In previous works \cite{Antoniadis:2002cs} more general models have been constructed that can lower the mass scale of the U(1) which can be more phenomenologically plausible but contain more moduli. It would be interesting to study minimal extensions of the ${\cal N }=0$ standard model including a massive U(1), as was done for the MSSM \cite{Anastasopoulos:2008jt}, to address a potential photon mass \cite{Reece:2018zvv}, relation to the the B-L symmetry \cite{Faraggi:1992rd}, the dark sector \cite{Mammarella:2012sx, Zhang:2022nnh} or other phenomenological implications \cite{Berezhiani:1996nu,Kurosawa:1999mn}. Other attempts include \cite{Avalos:2023ldc, Avalos:2023mti,Burgess:2003ic}.

Additionally, the rank of the three theories is even similar to the observations in supersymmetric theories \cite{Montero:2020icj} and in the  rigid theories \cite{Baykara:2023plc, Gkountoumis:2023fym}. Additionally, we noted that the rank of the 4d theory did not increase as a consequence of the quasicrystalline compactification while the 8d and 6d had the usual toroidal rank increase as those exist at special points of gauge symmetry enhancement.

As has been discussed throughout this work, string theory does give rise to non-supersymmetric vacua which can be related to supersymmetric vacua through a web of dualities \cite{Blum:1997gw,Blum:1997cs,Itoyama:1986ei,DeFreitas:2024ztt}. It would be interesting to further explore such dualities.

The theories we present are free of tachyons at tree level. However, there are two different one-loop effects that can give mass to tree level massless states. Firstly, the massless charged scalars can pick up a tachyonic/non-tachyonic mass at one-loop. This is the case for the $O(16)\times O(16)$ string on a circle $S^1$ at points with maximal gauge group enhancement \cite{Ginsparg:1986wr,Fraiman:2023cpa}. But also the 1-loop scalar potential discussed in \autoref{app:4d} will always add a   contribution to the charged under the U(1) scalars at 1-loop. The dilaton also acquires a mass term at 2-loops that is positive.

There is a way to cure these two instabilities for $O(16)\times O(16)$ and obtain a perturbatively stable non-supersymmetric AdS \cite{Baykara:2022cwj}. By tuning to weak coupling, the tachyons can be made light enough to be uplifted by the gravitational potential of AdS \cite{Breitenlohner:1982bm}, and by turning on flux, the dilaton gets stabilized. It would be worthwhile to attempt the same procedure for the new non-supersymmetric strings presented here. However, such constructions generally will have non-perturbative instabilities \cite{Ooguri:2016pdq, Fraiman:2023cpa}.

An interesting observation is that all leading order potentials are positive for our models as well as for the 10d non-susy tachyon-free theories. This seems to suggest that more ``rigid" tachyon free string theories always have positive cosmological constant. This is very reminiscent of our expectation from the real world! It would be very interesting to construct a counterexample or find a general argument why this has to be the case. There are potential examples with more moduli which have either negative cosmological constant \cite{Avalos:2023ldc} or vanishing cosmological constant at one loop \cite{Shiu:1998he,Kachru:1998yy, Kachru:1998hd}. 

A more in depth analysis is required for our models to determine their 1-loop fate in more general backgrounds. Especially in 4d, the potential term due to the anomalous $U(1)$ competes with the possibly tachyonic effects at 1-loop, which is interesting to study.

\textbf{Acknowledgments} 
HCT would like to thank Harvard University and the Swampland Initiative for their hospitality during part of this work. The work of ZKB and CV is supported by a grant from the Simons Foundation (602883,CV), the DellaPietra Foundation, and by the NSF grant PHY-2013858. 
\appendix
\section{Lattice constructions}\label{app:lattice}
\subsection{Gluing construction}
Gluing construction is used to obtain unimodular lattices from nonunimodular ones. It consists of combining the duals of lattices along their glue groups, as we now describe.

The \textit{glue group} of a lattice $\Lambda$ is given by
\begin{align}
    \mathcal D(\Lambda) = \Lambda^*/\Lambda.
\end{align}
An element of the glue group is an equivalence class denoted as $[v]$. The glue group inherits the norm of the lattice mod 2
\begin{align}
    |[v]|^2 := |v|^2 \pmod{2}.
\end{align}

Given two even lattices $\Lambda_1$ and $\Lambda_2$ with isometric glue groups
\begin{align}
    \psi: &\mathcal D(\Lambda_1)\to \mathcal D(\Lambda_2)\\
    &|v|^2=|\psi(v)|^2,
\end{align}
one can construct the unimodular lattice $\Lambda_1\Lambda_2$ as
\begin{align}
    \Lambda_1\Lambda_2 = \sum_{[v]\in\mathcal D(\Lambda_1)}\Lambda_1\oplus \Lambda_2' + (v,\psi(v)'),
\end{align}
in which the quadratic form is given by
\begin{align}
    |(a,b)|^2 = |a|^2 - |b|^2.
\end{align}
The primes for $\Lambda_2$ and $\psi(v)$ are there to remind us to flip the sign of their quadratic form.
\subsection{The $\mathbb Z_5$ quasicrystal}
As mentioned, $\mathbb Z_5$ rotational symmetry is not possible for 2d lattices, but are possible for 2d quasicrystals, which are slices of 4d lattices. We explain the formalism behind the construction of quasicrystals, and construct the $\mathbb Z_5$ quasicrystal in detail in \cite{Baykara:quasi}. Here, we give a quick construction of the indefinite lattice $\Gamma^{2;2}_5$ with the $(1/5,2/5)$ twist.

Take a starting basis vector $\alpha_0=(c_1,0;c_2,0)\in \mathbb R^{2;2}$. Then apply $\theta=R(2\pi /5)\oplus R(2\pi 2/5)$ to get $\alpha_i:=\theta^i\cdot \alpha_0$ for $i=1,\dots,3$. Choose $c_1$ and $c_2$ so that the determinant is minimal and the lattice is even. Then the lattice $\Gamma^{2;2}_5$ is generated by
\begin{align}
    \alpha_0 &= \frac{\sqrt{2}}{\sqrt[4]{5}}\left(1,0;1,0\right),\\
    \alpha_1 &= \frac{\sqrt{2}}{\sqrt[4]{5}}\left(\frac{\sqrt{5}-1}{4},\frac{\sqrt{5+\sqrt{5}}}{2\sqrt{2}};\frac{-\sqrt{5}-1}{4},\frac{\sqrt{5-\sqrt{5}}}{2\sqrt{2}}\right),\\
    \alpha_2 &= \frac{\sqrt{2}}{\sqrt[4]{5}}\left(\frac{-\sqrt{5}-1}{4},\frac{\sqrt{5-\sqrt{5}}}{2\sqrt{2}};\frac{\sqrt{5}-1}{4},-\frac{\sqrt{5+\sqrt{5}}}{2\sqrt{2}}\right),\\
    \alpha_3 &= \frac{\sqrt{2}}{\sqrt[4]{5}}\left(\frac{-\sqrt{5}-1}{4},-\frac{\sqrt{5-\sqrt{5}}}{2\sqrt{2}};\frac{\sqrt{5}-1}{4},\frac{\sqrt{5+\sqrt{5}}}{2\sqrt{2}}\right).
\end{align}
The determinant is $5$, so it is not unimodular.

By the gluing construction, $\Gamma^{2;2}_5\Gamma^{2;2}_5$ is unimodular. More explicitly, one takes two copies of $\Gamma^{2;2}_5$, flip the left and right movers for one of them, and glue along two copies of the glue vector
\begin{align}
    w&=-\frac 2 5 \alpha_0 + \frac 1 5 \alpha_1 -\frac 1 5 \alpha_2 +\frac 2 5 \alpha_3\\ \label{eq:wnorm}
    |w|^2&=\frac 2 5
\end{align}
as
\begin{align}
    \Gamma^{2;2}_5 \Gamma^{2;2}_5 = \sum_{n=0}^4 \Gamma^{2;2}_5 \oplus \Gamma^{2;2'}_5+ n(w,w'),
\end{align}
where the left and right movers for the primed lattice $\Gamma^{2;2'}_5$ and the glue vector $w'$ are switched compared to the unprimed ones. The result is a unimodular lattice with twists $(0,1/5;0,2/5)$ and $(1/5,0;2/5,0)$.

\subsection{Lattice of dimension 8 and determinant 5}
Given that the determinant of the $\mathbb Z_5$ quasicrystal is $5$, it is possible to construct a unimodular lattice by gluing it to another lattice $L$ of dimension $8$ and determinant $5$. In \cite{Conway1988LowdimensionalLI}, it was shown that there is a unique even lattice in dimension $8$ with determinant $5$, whose quadratic form is given by the Gram matrix
\begin{align}
    G_L=\left(
\begin{array}{cccccccc}
 2 & -1 & 0 & 0 & 0 & 0 & 0 & 0 \\
 -1 & 2 & -1 & 0 & 0 & 0 & 0 & 0 \\
 0 & -1 & 2 & -1 & 0 & 0 & -1 & 0 \\
 0 & 0 & -1 & 2 & -1 & 0 & 0 & 0 \\
 0 & 0 & 0 & -1 & 2 & -1 & 0 & 0 \\
 0 & 0 & 0 & 0 & -1 & 2 & 0 & 0 \\
 0 & 0 & -1 & 0 & 0 & 0 & 2 & 1 \\
 0 & 0 & 0 & 0 & 0 & 0 & 1 & 6 \\
\end{array}
\right).
\end{align}
The lattice $L$ can be thought of as the root lattice $\Gamma(E_7)$ together with a long vector $\alpha_8$ with $\alpha_8^2=6$ and $\alpha_7\cdot \alpha_8=1$, where $\alpha_i$ for $i=1,\dots,7$ are simple roots of $E_7$.

The glue vector of $L$ is given by
\begin{align}
    b&=\frac 1 5(\alpha_1+2\alpha_2-2\alpha_3+\alpha_4-\alpha_5+2\alpha_6-2\alpha_7+2\alpha_8)\\
    |b|^2&=\frac 2 5.
\end{align}
Comparing with \eqref{eq:wnorm}, we see that the glue vectors of $\Gamma^{2;2}_5$ and $L$ are isometric. Therefore, we can use the gluing construction to get the unimodular lattice
\begin{align}
    L\Gamma^{2;2}_5 = \sum_{n=0}^4 L\oplus \Gamma^{2;2'} + n(b,w').
\end{align}
The lattice has the twist $(1/5,0^4;2/5)$.

\section{1-loop potential}\label{app:constant}
The one-loop cosmological constant is obtained by integrating the partition function over the fundamental domain $\mathcal F$ of the torus as
\begin{equation}
    \Lambda_{\text{1-loop}}=-{1\over (2\pi\sqrt{\alpha'})^d}\int_{F}{d^2\tau\over 2\tau_2^{d/2+1}}Z(\tau ).
\end{equation}
We present the partition function of each $\hat g^n$-twisted sector, and compute the cosmological constant. Similar to the $O(16)\times O(16)$ string, all of them have positive cosmological constant.

\subsection*{4d}
The partition function $Z_n$ of the $\hat{g}^n$-twisted sector is
\begin{align}
\begin{split}
    Z_0 &= 2 q^{-1}+ 30q^{-1}\bar q-10+150 \bar q -158 q \\
    &\qquad +330 q \bar q+\dots
\end{split}\\
    \begin{split}
        Z_1=Z_4 &= -215+4125 q^{1/5} \bar q^{1/5} -33250q^{2/5}\bar q^{2/5}\\&\qquad +126375 q^{3/5}\bar q^{3/5}-2150\bar q\\&\qquad-239175(1+10\bar q) q+\dots
    \end{split}\\
    \begin{split}
        Z_2=Z_3 &= 90-8000q^{2/5}\bar q^{2/5}-30125 q^{3/5}\bar q^{3/5}\\&\quad +19250 q^{4/5}\bar q^{-1/5}+115500 q^{4/5}\bar q^{4/5} \\&\qquad+ 675 \bar q+113650q+852375 q\bar q+\dots
    \end{split}
\end{align}
The cosmological constant at one loop is
\begin{align}
    -\frac{1}{(2\pi \sqrt{\alpha'})^4}\int_{\mathcal F} \frac{1}{2\tau_2^3}Z(\tau) \approx \left(3.13 \times 10^{-2}\right)\alpha'^{-2}.
\end{align}

\subsection*{6d}
The partition functions are given as
\begin{align}
    Z_0 &= 4 q^{-1}+20q^{-1}\bar q-4-4 q-20\bar q -20 q \bar q+\dots\\
    \begin{split}
        Z_1&=Z_2=Z_3=Z_4 
    \\&= -125-625 q^{1/5}\bar q^{1/5}+142500q^{4/5}\bar q^{4/5}-625\bar q\\&+23750 q^{4/5} \bar q^{-1/5}-63125 q -315625 q\bar q+\dots
    \end{split}
\end{align}
The cosmological constant at one loop is
\begin{align}
    -\frac{1}{(2\pi \sqrt{\alpha'})^6}\int_{\mathcal F} \frac{1}{2\tau_2^4}Z(\tau) \approx \left(2.89 \times 10^{-3}\right)\alpha'^{-3}.
\end{align}

\subsection*{8d}
The partition functions are given as
\begin{align}
    Z_0 &= 6q^{-1}+54 q^{-1}\bar q -18-162 \bar q+\dots\\
    \begin{split}
        Z_1=Z_2&= -729 +13122 q^{2/3}\bar q^{-1/3}+183708 q^{2/3}\bar q^{2/3}\\ &\quad -6561 \bar q-183708 q-1653370 q \bar q+\dots
    \end{split}
\end{align}
The cosmological constant is
\begin{align}
    -\frac{1}{(2\pi \sqrt{\alpha'})^8}\int_{\mathcal F} \frac{1}{2\tau_2^5}Z(\tau) \approx \left(1.26 \times 10^{-4}\right)\alpha'^{-4}.
\end{align}

\section{Anomalies}\label{app:anomalies}
In this appendix we will reviews general aspects of anomalies in 4,6,8 dimensions used to demonstrate the consistency of our models. Of course in string theory anomalies are cancelled automatically and specifically in the heterotic string theory both the anomaly term and the Green-Schwarz term come from 1-loop $1+D/2$ amplitudes.

\subsection{4d Anomalies}\label{app:4d}

 In 4d CP acts by exchanging chirality and hence terms odd under charge conjugation cancel automatically. Anomalies are captured by triangle diagrams with external gauge fields /gravitons and internal chiral fermions.  

All irreducible anomalies are expected to cancel automatically as they do not participate in the Green-Schwarz mechanism \cite{Green:1984sg} which we will discuss below. 

The condition for the non-abelian gauge anomaly cancellation is
$    \sum_i E_{R_i}^G=0
$
where $\Tr_{\cal R}(F^3)=E_{\cal R}tr_F(F^3)$
For $SO(10)$ its trivially satisfied since all representations are real. For the other gauge groups we have 
\begin{eqnarray}
E_{5/10}^{SU(5)}=E_{3}^{SU(3)}=E_{2}^{SU(2)}=1
\end{eqnarray}
and its simple check that all anomalies cancel.
The abelian factors can be more subtle and they could have anomalous $U(1)-\text{Graviton}-\text{Graviton}$ diagrams exactly when $\Tr(Q)\neq 0$. Such an anomaly was discussed in \cite{Alvarez-Gaume:1983ihn} and can be cancelled by the  Green-Schwarz mechanism\cite{Green:1984sg}. Such U(1)s are called ``anomalous" and in fact  they will become massive. 

Consider the situation of one ``anomalous" $U(1)_X$ which applies to the heterotic string theory with a single universal axion and hence our model. Then the gauge transformation $A^X_\mu \to A^X_\mu +\partial_\mu \epsilon $ leads to the anomalous variation 
\begin{align}
\begin{split}
    \delta_X S=-&{1\over 32\pi^2}\int  \epsilon\{c_XF_{U(1)_X}^2+\sum_ac_a\Tr(F_{G_a}^2)-c_{\cal R}\Tr(R^2)\\
   & +\sum_ic_{Xi}F_{U(1)_X}\wedge F_{U(1)_i}+\sum_{ij}c_{ij}F_{U(1)_i}\wedge F_{U(1)_j})\}
   \end{split}
\end{align}
where $c_X={1\over 3}\Tr(Q^3_X),c_a={\Tr(A^G_{{\cal R}_i}Q_X)}$ $ c_{\cal R}={1\over 24}\Tr(Q_X), c_{Xi}=2\Tr(Q_X^2Q_i), c_i=2\Tr(Q_XQ_i^2)$ with $ Tr(F^2)_R=A_{\cal R}^G Tr_F(F^2)$ .

We use the Green-Schwarz mechanism \cite{Green:1984sg} to cancel these anomalies. In 4d the $B_{\mu \nu}$ field is Poincare dual to a scalar axion $\theta$ and the axion participates in the anomaly cancellation \cite{Witten:1984dg} by a transforming \cite{Harvey:1988in} as $\delta_\epsilon \theta= {1\over 64\pi^2}c_X M\epsilon$ for some mass $M$ which is identified with $M_s$ in string theory. So we can equivalently describe the GS mechanism using the axion.

\begin{align}\label{eq:4dGS}
\begin{split}
    \mathcal{L}_\theta &=-\sum_i{1\over 4 g_{U(1)_i}^2}|F_{U(1)_i}|^2-\sum_a{1\over 4 g_{G_a}^2}| F_{G_a}|^2\\ & \  -(\partial_\mu \theta +{1\over 64\pi^2}c_X M\ A_\mu^X)^2 \\&  +{2\theta \over M c_X } \{\sum_{ij}c_{ij} F_{U(1)_i}\wedge F_{U(1)_j}+ \sum_a c_a \Tr(F_{G_a}^2)\\ &  +\sum_{i}c_{Xi} F_{U(1)_X}\wedge F_{U(1)_j}+ \sum_{X}c_{X} F_{U(1)_X}^2-c_R\Tr(R^2)\}\\&+\sum_iA_{U(1)_i}\wedge A_{X}\wedge (c_{Xi}F_X+c_{ij}F_{U(1)_i})
    \end{split}
\end{align}
the last term represents the \textit{generalized Chern-Simons} term \cite{Anastasopoulos:2006cz} which is needed to preserve the gauge invariance of the non-anomalous U(1)s

 Note that the Green-Scwarz mechanism we have presented implies that the anomalous U(1)  gains a St\"uckelberg  mass $m_X={1\over 64\pi^2}c_X M$.Then the axion is ``eaten" by the anomalous U(1) and serves as the longitudinal component of the massive vector. Therefore, the only real neutral scalar field that remains is the dilaton.

The various group theory constants in \autoref{eq:4dGS} satisfy the relation
\begin{align}\label{4danomaly}
  \Tr_{G_a}(A_{R_i}^{G
}Q_X)=\Tr(Q^2_jQ_X)={1\over 3}\Tr( Q_X^3)={1\over 24}\Tr(Q_X) \ 
\end{align}
where the group generators are all normalized to have level 1.
A simple check will show that indeed our model satisfies these \autoref{4danomaly}. As we showed above the mass of the massive U(1) needs to be   ${1\over 64 \pi^2}\Tr(Q^3)M$ to cancel the anomalies. However, in $\mathcal{N}=1$ 4d string compactifications it has been shown that $M_{X}={1\over 192 \pi^2}\Tr(Q)M_s$ \cite{Dine:1987gj,Atick:1987gy} for the universal axion GS which supports \autoref{4danomaly}\footnote{M of \autoref{eq:4dGS} represents a UV mass scale. In string theory this scale is given by $M=M_s$.}. More, generally for any pertubative compactification of the heterotic string, worldsheet modular invariance has been shown \cite{Schellekens:1986yj,Schellekens:1986xh} to imply the relations \autoref{4danomaly} which are indeed satisfied for our model. 
The mass of the ``anomalous" U(1) has been computed in ${\cal N }=1,2$ type I string theory \cite{Antoniadis:2002cs}
and non-supersymmetric Type I in \cite{Anastasopoulos:2004ga} with more general axion couplings.

In our case the axion corresponds to the dual of the Kalb-Ramond field. The coupling with the massive U(1) is given by $M_XB\wedge F_X$ as discussed above and hence the coupling to the axion $M_X\partial_\mu \theta A^\mu $. The mass of the U(1) comes from the two point amplitude of $A_\mu^X$ with an internal $\theta$ and hence the mass term of the U(1) is given by $M^2_X(A_\mu^X)^2$. In string theory this term comes from a 2-loop diagram \cite{Dine:1987xk} and hence the mass term in Einstein frame is given by $e^{6\phi}M^2_{X}$. Therefore, the U(1) gets massed up and ``eats" the axion to become its longitudinal mode.

In the $\mathcal{N}=1$ case a massive U(1) has been shown to give rise to a Fayet-Illiopoulos term \cite{
Dine:1987gj,Atick:1987gy} at 1-loop which gives mass to the charged scalars while the dilaton picks up a mass at 2-loops \cite{Atick:1987qy}. The potential in this case is given by 
\begin{eqnarray}
    V_D=e^{-2\sqrt{2}\hat{\phi}}g^2_s\left(e^{\sqrt{2}\hat{\phi}}M_X^2+\sum_i q_i\varphi^*_i\varphi_i\right)
\end{eqnarray}
in terms of the canonically normalized dilaton.

Additionally, we can also compute the 1-loop beta function for our model. For a 4d gauge theory with gauge group $G=\prod G_i$, the 1-loop beta function $\beta(g_i)$ corresponding to $G_i$ is given by  
\begin{align}
    \beta(g_i)&=\frac{g_i^3}{16\pi^2} b_i\\
    b_i&=-{11\over 3}I(R_{adj})+{1\over 3}\sum_{\substack{\text{Complex}\\\text{scalars}}} I(R_s)+{2\over 3}\sum_{\substack{\text{Weyl}\\\text{fermions}}}I(R_f)
\end{align}
where $I$ corresponds to the index of a representation e.g. for $SU(N)$ we have $ I(F)={1\over 2}, I(R_{adj})=N$. For abelian factors $G_i=U(1)$, we have $I(R_{adj})=0$ and $I(q)=q^2$, where $q$ is the $U(1)$ charge. We find
\begin{align}
    b_{SO(10)}&= -9\\
    b_{SU(5)} &= -\frac 3 2\\
    b_{SU(3)} &= \frac{67}{2}\\
    b_{SU(2)} &= \frac{121} 6\\
    b_{U(1)} &=\left(1610,4070,\frac{9250}{3},34440\right).
\end{align}

The order of the $U(1)$ factors is the same as in \autoref{tab:nonsusy4d}. We see that only the $SO(10)$ and $SU(5)$ factors are asymptotically free.

\subsection{6d Anomalies}
The 6d local anomalies are computed in \cite{Alvarez-Gaume:1983ihn}. In our case we have one single 2-form field which splits into one self-dual and one anti self-dual two form and hence they do not contribute to the anomalies. Therefore, the 8-form anomaly  receives contributions from only chiral fermions of both chiralities since in 6d they are independent. 

\begin{align}
    I_8=\sum_i(  n_{+}^{\mathcal{R}_i}  \sum_a I_{1/2}(F^a_{\mathcal{R}_i},R)+ n_{-}^{\mathcal{R}_i}  \sum_a I_{1/2}(F^a_{\mathcal{R}_i},R))
\end{align}
The Green-Schwarz mechanism\cite{Green:1984sg} requires the anomaly polynomial to be reducible and split in the from 

\begin{eqnarray}
 (2\pi)^4  I_{8}={1\over 12}X_4Y_4
\end{eqnarray}
 Anomaly cancellation is achieved by introducing the counterterm $B\wedge X_4$.

Note that the irreducible terms $Tr(R^4),Tr(F^4)$ need to cancel identically which gives the conditions 
\begin{align}
  Tr(F^4)&:  \sum_{ia} (n^{\mathcal{R}_i^a}_+-n^{\mathcal{R}_i^a}_- ) B_{\mathcal{R}^a_i}=0\\
  Tr(R^4)&: \sum_a (n^a_+-n^a_-)=0\\
  Tr(F^3_a)F_b&:  \sum_{ia} (n^{\mathcal{R}_i^{ab}}_+-n^{\mathcal{R}_i^{ab}}_- ) E_{\mathcal{R}^a_i}Q_b=0
\end{align}
where the group theory invariants are defined in \ref{groupinvariants} with $A_{5}^{SU(5)}=1,A_{10}^{SU(5)}=3,E_{5/10}^{SU(5)}=1$.
Note that $Tr(F)$ is automatically zero for both non-abelian and abelian theories because in the former $\Tr(T^a)=0$ and for the latter CP conjugates in 6d have the same chirality but conjugate representations  $q\to-q, R\to \bar{R}$.

\subsection{8d Anomalies}\label{app:8d}
In this case anomaly contributions come from one-loop diagrams with five external legs of gauge fields or gravitons and internal chiral fermions. Since CP conjugation exchanges chirality we will only consider states of definite chirality.

The total anomaly polynomial consists of anomalies that are reducible and irreducible. The reducible term are such that 
\begin{eqnarray}
  (2\pi)^5  I_{10}^{red.}={1\over 12}X_6\wedge Y_4
\end{eqnarray}

which means that they can be cancelled by a Green-Schwarz counter term 
\begin{eqnarray}
    B\wedge X_6
\end{eqnarray}
with the appropriate transformation properties of $\delta B$ that cancel the anomalies as described in \cite{Green:1984sg}.
The irreducible anomalies (e.g. $F_iR^4$) need to cancel identically since they do not participate in GS and hence we do not describe them but they are satisfied for our models.
We note that anomalies coefficients  with representations invariant under  charge conjugation will be automatically zero like in 4d because of CPT invariance.

The reducible anomalies are given by $\tilde{I}_{10}^{red.}= 12 (2\pi)^3I_{10}^{red.}$

\begin{align}
\begin{split}
&  I_{10}^{red.}={1\over 10}\sum_ia_i\Tr(F^5_i)+{1\over 10}\sum_{m}c_{m}\Tr(F_m^3)^2\\& +\sum_{I\neq J}c_{IJ}\Tr(F_I^2)\Tr(F_J^3)   +{1\over 2}\sum_{i\neq J}b_{iJ}\Tr(F_i)\Tr(F^4_J)\\&+{1\over 2}\sum_{i,m}f_{im}\Tr(F_i)\Tr(F_m^2)^2+3\sum_{i,J,K}f_{iJK}\Tr(F_i)\Tr(F_J^2)\Tr(F_K^2) \\& +{1\over 24}\sum_Id_I\Tr(F_I^3)\Tr(R^2)+{1\over 8}\sum_{i,J}e_{iJ}\Tr(F_i)\Tr(F_J^2)\Tr(R^2)
\end{split}
\end{align}

where $i$ is over the two abelian groups and $I={i,m}$ over all gauge groups and the various group theoretic  constants are given by  $a_i=\Tr(Q^5_i), b_{ij}=\Tr(Q_iQ^4_j),b_{im}=\Tr(Q_iB_{\mathcal{R}_m}), c_{ij}=\Tr(Q^2_iQ^2_j),c_{im}=\Tr(Q^2_iE_{\mathcal{R}_m}), c_{mn}=\Tr(A_{\mathcal{R}_m}E_{\mathcal{R}_n}),c_m=\Tr(K_{\mathcal{R}_m}),f_{ijm}=\Tr(Q_iQ_j^2A_{\mathcal{R}_m}),f_{inm}=\Tr(Q_iA_{\mathcal{R}_n}A_{\mathcal{R}_m}),f_{im}=\Tr(Q_iC_{\mathcal{R}_i}),d_i=\Tr(Q_i^3),d_m=\Tr(E_{\mathcal{R}_m}), e_{ij}=\Tr(Q_iQ^2_J),e_{im}=\Tr(Q_iA_{\mathcal{R}_m})$

 with the following relations
\begin{align}    \label{groupinvariants}
\begin{split}
&\Tr_\mathcal{R}(F^2)=A_\mathcal{R}\Tr_V(F),\Tr_\mathcal{R}(F^3)=E_\mathcal{R}\Tr_V(F^3)\\
&\Tr_\mathcal{R}(F^4)=B_\mathcal{R}\Tr_V(F^4)+C_\mathcal{R}\Tr_V(F^2)^2\\& \Tr_\mathcal{R}(F^5)=D_\mathcal{R}\Tr_V(F^5)+K_\mathcal{R}\Tr_V(F^3)\Tr_V(F^2)
\end{split}
\end{align}
with 
$D_{84}=-27,\ D_{9}=1,E_{84}=9,\ E_{9}=1,A_{84}=21,A_{9}=1,C_{84}=15,C_9=0, B_{89}=-9,B_9=1,K_{84}=30$.

We note that consistency requires $12 \Tr(Q^5)=5\Tr(Q^3)$ and $\Tr(Q)=0$ for the U(1) factors to factorize.

There could also be anomalies similar to those in \ref{app:4d} for U(1)s but are not present in this model and hence we do not discuss them here.

\let\bbb\bibitem\def\bibitem{\itemsep4pt\bbb}
\bibliography{refs}

\end{document}